\documentclass[a4paper]{jpconf}
\usepackage{graphicx}
\usepackage{hyperref}
\usepackage{color}
\begin{document}
\title{Event reweighting with the NuWro neutrino interaction generator}

\author{Luke Pickering$^{\,1}$, Patrick Stowell$^{\,2}$, Jan Sobczyk$^{\,3}$}
\address{$^{1\,}$Imperial College London, $^{2\,}$University of Sheffield, $^{3\,}$University of Wroc\l{}aw}

\ead{$^{1\,}$lp208@ic.ac.uk}

\newcommand{\figpreref}{{Figure.~}}
\newcommand{\tablepreref}{{Table.~}}
\newcommand{\caf}{\textit{C}^\textsc{\tiny A}_\textsc{\tiny 5}\hspace{-1mm}\left(0\right)}
\newcommand{\mares}{\textit{M}^\textsc{\tiny A}_\textsc{\tiny RES}}
\newcommand{\maqe}{\textit{M}^\textsc{\tiny A}_\textsc{\tiny QE}}
\newcommand{\qsq}{Q^{2}}
\newcommand{\pip}{\pi^{+}}
\newcommand{\senu}{{\it\sigma}\left(E_{\nu}\right)}
\newcommand{\tcaf}{$\caf$}
\newcommand{\tmares}{$\mares$}
\newcommand{\tmaqe}{$\maqe$}
\newcommand{\tqsq}{$\qsq$}
\newcommand{\tpip}{$\pip$}
\newcommand{\tsenu}{$\senu$}
\newcommand{\tnrbkg}{$\textit{NR}^\textsc{\tiny BKG}$}

\begin{abstract}

Event reweighting has been implemented in the NuWro neutrino event generator for a number of free theory parameters in the interaction model.
Event reweighting is a key analysis technique, used to efficiently study the effect of neutrino interaction model uncertainties.
This opens up the possibility for NuWro to be used as a primary event generator by experimental analysis groups.
A preliminary model tuning to ANL and BNL data of quasi-elastic and single pion production events was performed to validate the reweighting engine.

\end{abstract}

\section{NuWro}

NuWro \cite{nuwro_12} is a neutrino interaction generator capable of producing predictions for neutrino-nucleus interactions at neutrinos of energies between 0.1 and 100 GeV.
NuWro contains a wide variety of models and tuneable parameters, but until now it has had no facility to perform event reweighting.
This has limited its use by experimental groups for anything more than final cross-section prediction comparisons. 
NuWro ReWeight was written to facilitate the use of NuWro at current and future neutrino interaction experiments.

\section{Event Reweighting}

\begin{figure}[h]
\begin{minipage}{0.45\textwidth}
\begin{center}	
\includegraphics[width=0.95\textwidth]{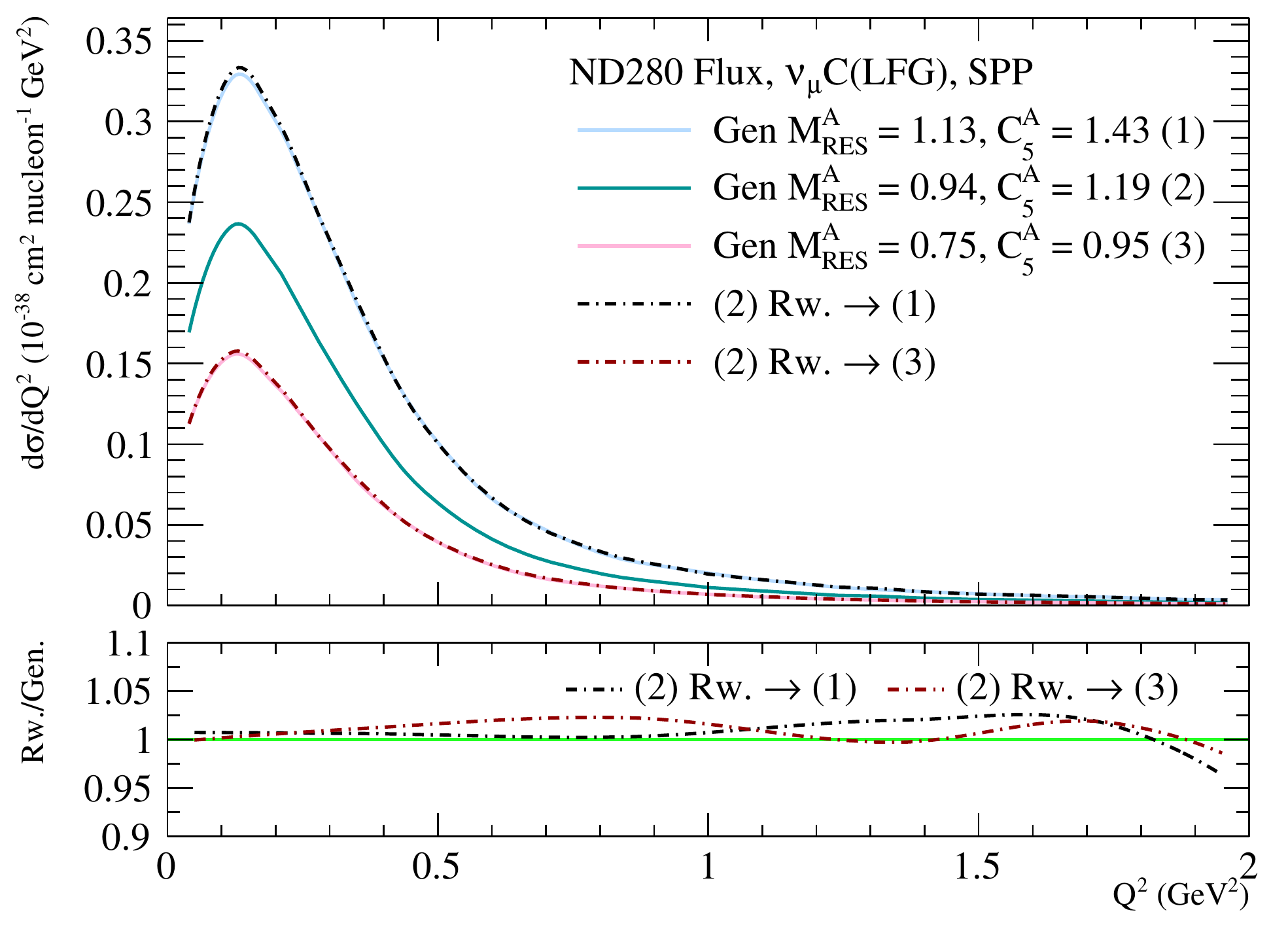}
\caption{\label{fig:rwvalid} Solid lines demarcate generated events. Dashed lines show the results of reweighting generated event set (2) to use the values of \tmares\ and \tcaf\ from (1) and (3).}
\end{center}
\end{minipage}\hspace{2pc}%
\begin{minipage}{0.45\textwidth}
\begin{center}
\includegraphics[width=0.95\textwidth]{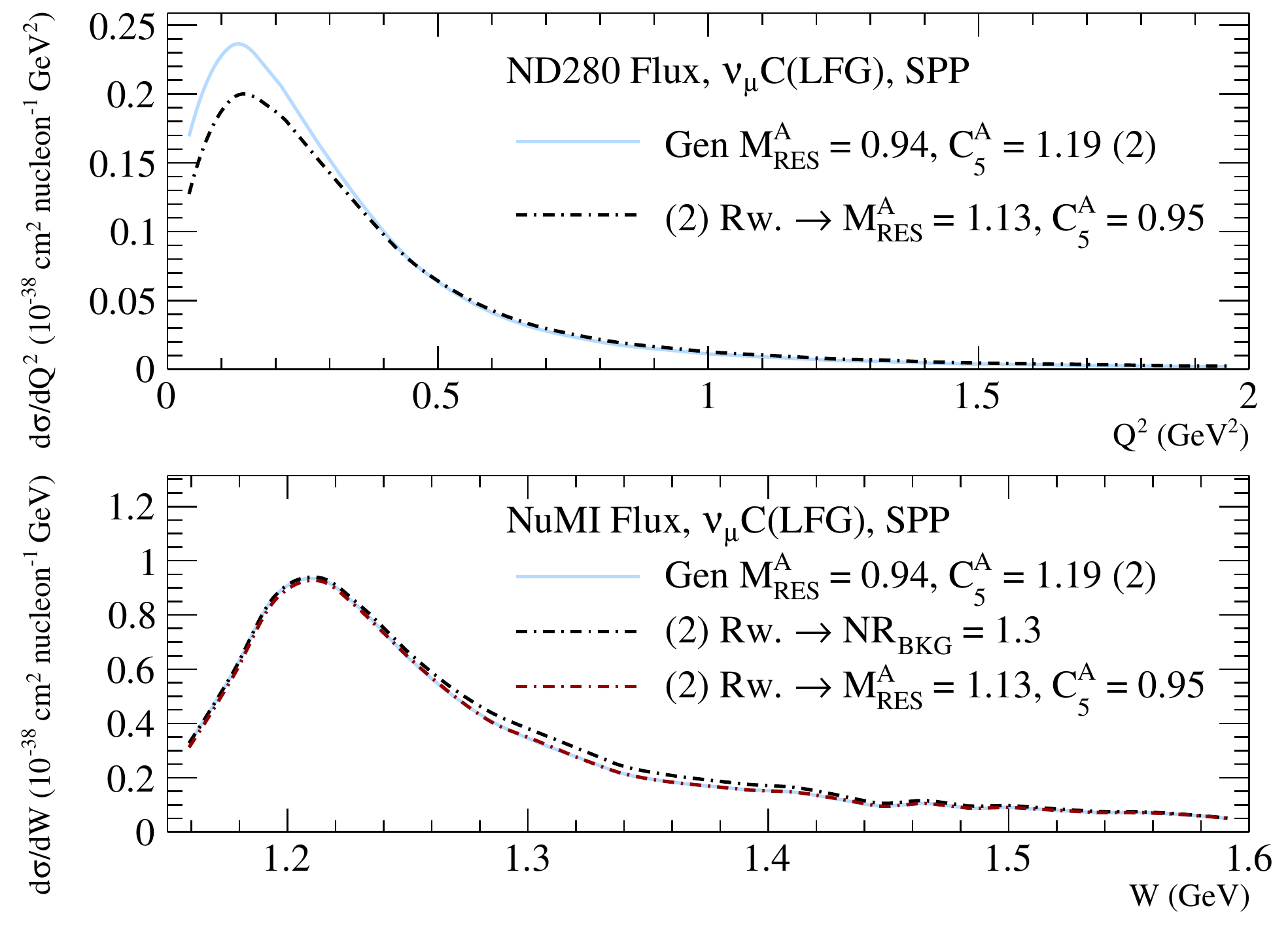}
\caption{\label{fig:dialresponse} The response of \tqsq\ and $W$ to some variations of the SPP parameters. \tmares\ and \tcaf\ have similar effect over most of the \tqsq\ range. Increasing \tnrbkg\ fills in the transition region between Delta resonance production (1232\,GeV) and DIS.}
\end{center}
\end{minipage} 
\end{figure}

Event generation is an inherently inefficient process.
Generated event properties must be distributed correctly according to the model used.
This is often achieved by rejection sampling: randomly throwing sets of interaction properties and accepting interactions with a probability proportional to the predicted cross-section for that event.
When tuning free model parameters to data, many sets of model predictions must be generated to investigate how the model fits the data.
It is often advantageous to `reweight' model predictions to determine the response of varying free parameters rather than fully recalculating them.
Each generated event comes with an associated `weight' which is proportional to the probability for that event.
Reweighting is the process of calculating a scaling factor which can be combined with the original event weight to give a new weight which would be correct if the event had been generated under some different set of free parameter values.
This process involves no Monte Carlo techniques and so no computational time is `wasted', it can be many orders of magnitude faster than re-generation.
This significant boost in efficiency makes more involved studies of systematic uncertainty and goodness-of-fit feasible. 
Generating $5\times10^{5}$ events using the NuMI on-axis flux was found to take $\mathcal{O}\left(500\,s\right)$, while reweighting those events to a new prediction took $\mathcal{O}\left(3\,s\right)$ on a single CPU core, and $\mathcal{O}\left(0.2\,s\right)$ when parallelised over 32 cores.

NuWro now supports reweighting of free-nucleon model parameters for Charged Current Quasi Elastic (CCQE) \cite{lsaccelreact_72} and single pion production (SPP) interactions---important interaction channels at beam energies of $\mathcal{O}\left(1\,\textrm{GeV}\right)$.
Three free model parameters are available for SPP events:
\tcaf\ and \tmares\ are free parameters in the baryonic resonance form factors\,\cite{GraczykSobczykFF_08}.
The other free parameter, \tnrbkg\ is a scale factor for the cross section of SPP through non-resonant processes.
\figpreref\ref{fig:rwvalid} shows the effect of simultaneously varying \tmares\ and \tcaf\ on the SPP \tqsq\ distribution.
It can be seen that the reweighting very precisely reproduces the fully re-generated prediction.
\figpreref\ref{fig:dialresponse} shows the effect of reweighting the three parameters on the SPP cross section as a function of $Q^2$ and $\textrm{W}$.
Increasing the value of \tcaf\ and decreasing \tmares\ simultaneously results in a cross section prediction that is very similar to the nominal prediction over a large fraction of the $Q^2$ range; this suggests that these parameters are strongly anti-correlated.

\section{Comparison to Bubble Chamber data}

\begin{figure}[h]
\begin{minipage}{0.45\textwidth}
\includegraphics[width=\textwidth]{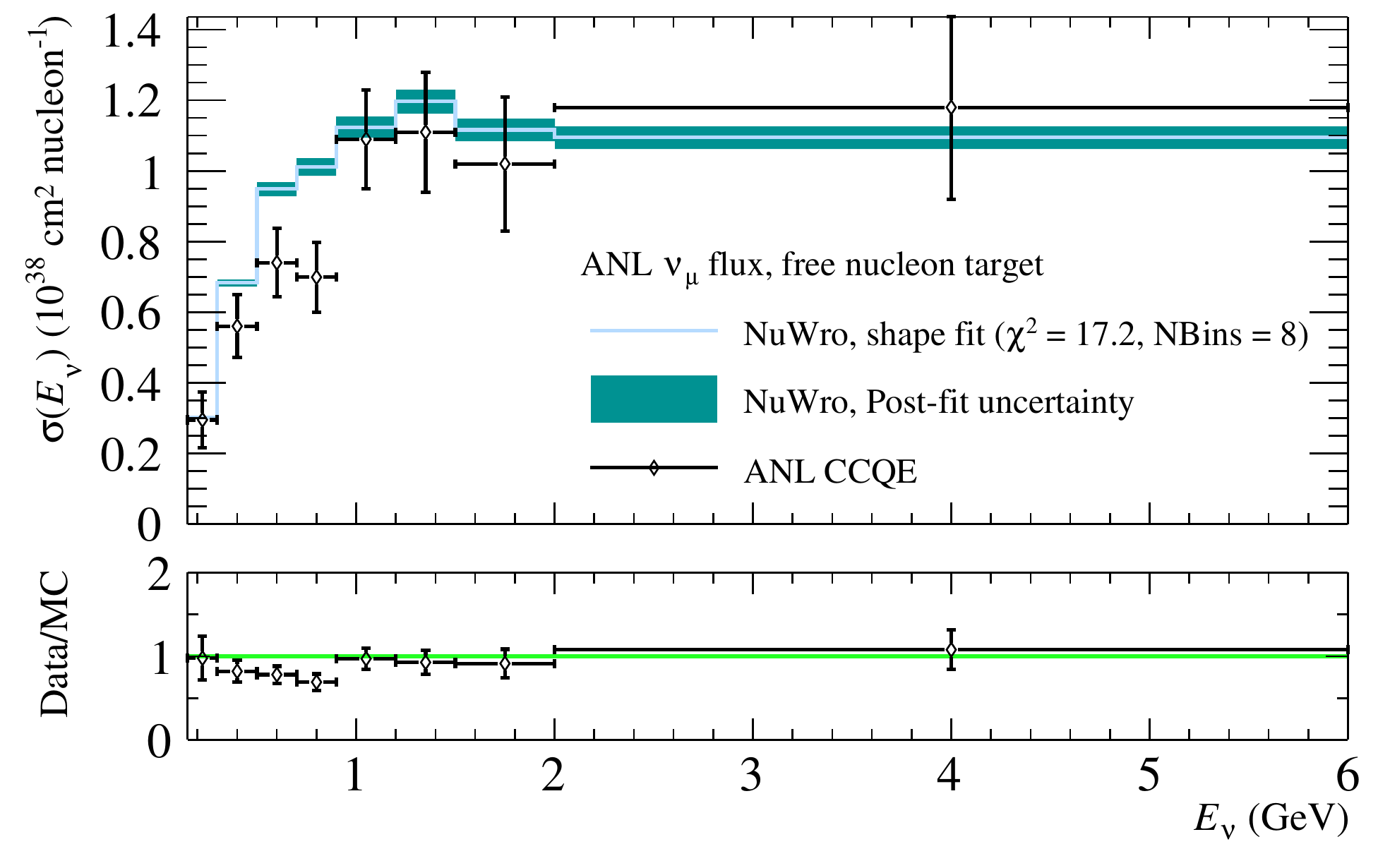}
\caption{\label{fig:anlccqeenu}Example post-fit comparison to ANL CCQE \tsenu distribution.}
\end{minipage}\hspace{2pc}%
\begin{minipage}{0.45\textwidth}
\includegraphics[width=\textwidth]{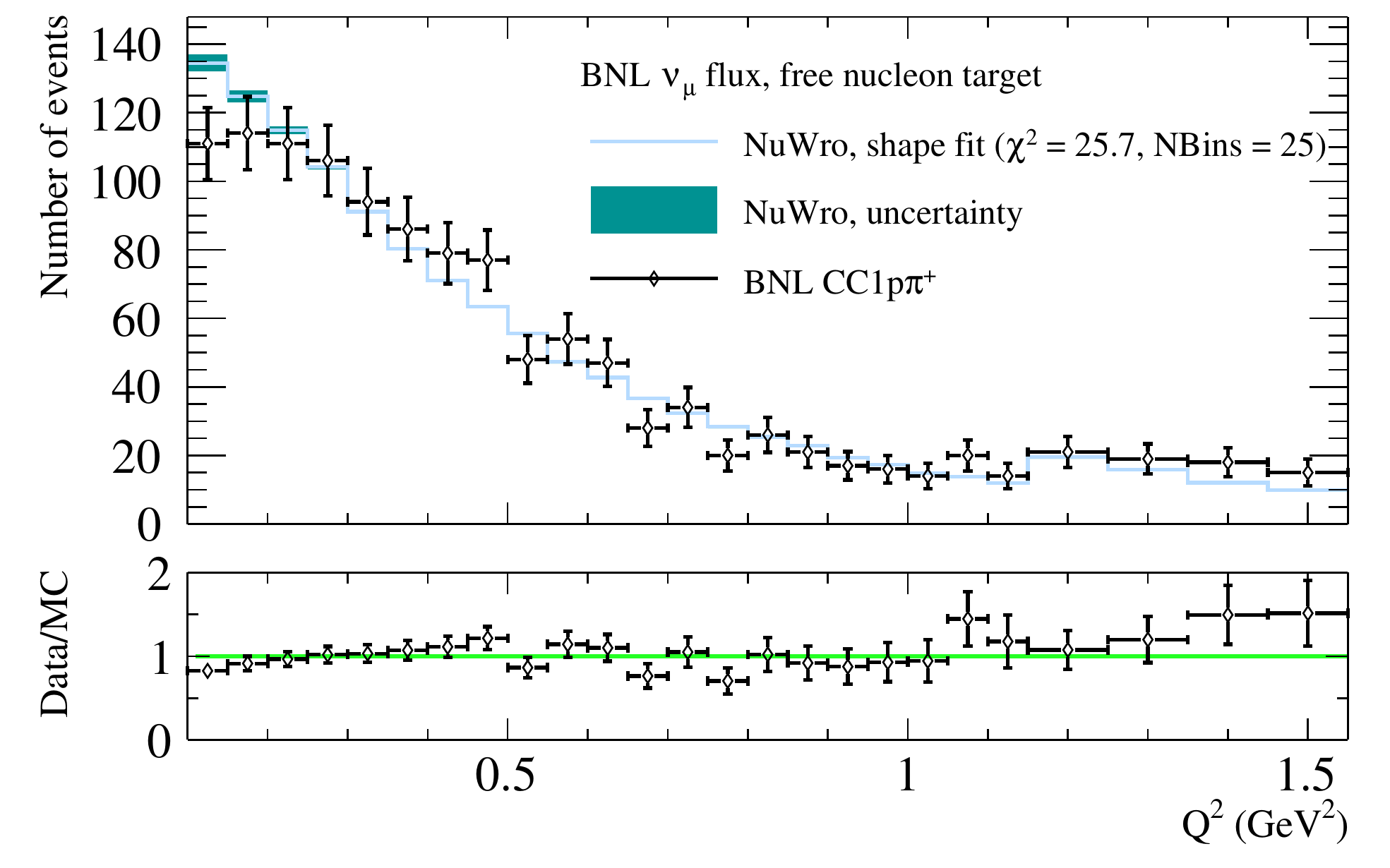}
\caption{\label{fig:bnlcc1pipq2}Example post-fit comparison to BNL CC1\tpip \tqsq event rate distribution.}
\end{minipage} 
\end{figure}

\begin{figure}[h]
\begin{minipage}{0.45\textwidth}
\begin{tabular}{@{}l*{15}{l}{l}}
\br
Param & Nominal & Best fit\\
\mr
\tmaqe/GeV & $1.20$ & $1.05 \pm 0.03$\\
\tmares/GeV & 0.94 & $0.93 \pm 0.03$\\
\tcaf & 1.19 & $0.94 \pm 0.03$\\
\tnrbkg & 0.00 & $1.35 \pm 0.13$\\
\br
\end{tabular}
\caption{\label{table:ANLBNLFitResults}Free-nucleon fit results for the four free  parameters in the CCQE and single pion production interaction models.}
\end{minipage}\hspace{2pc}%
\begin{minipage}{0.45\textwidth}
\includegraphics[width=0.8\textwidth]{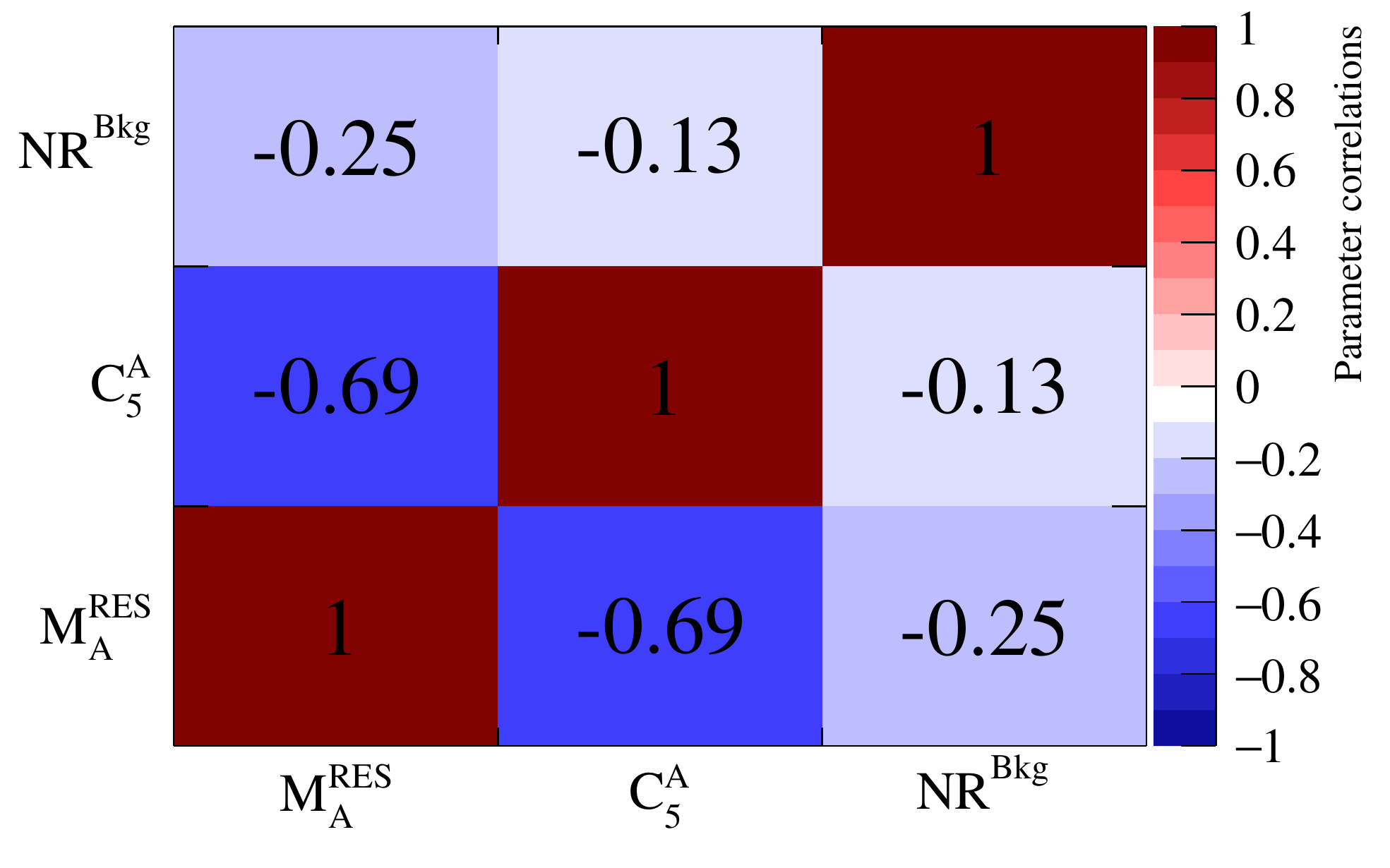}
\caption{\label{fig:covmat}The post-fit correlation matrix for the SPP free parameters. \tmares\ and \tcaf\ show a high degree of anti-correlation, which is to be expected.}
\end{minipage} 
\end{figure}

To test the event reweighting, as well as the current predictions of NuWro, reweighting was used to tune the NuWro CCQE and SPP predictions to historic bubble chamber data from ANL~\cite{ANLCCQE_77,ANL1pi_82} and BNL~\cite{BNLCCQE_81,BNL1pi_86}.
Comparison to bubble chamber data is important because interactions on deuteron targets are expected to only exhibit weak final state interaction effects \cite{wu_15}. This allows tuning of the neutrino--nucleon interaction model. 
Subsequent comparison to nuclear-target data can be used to tune theoretical models of nuclear effects.

The global neutrino cross-section comparison framework, NUISANCE \cite{NUISANCE}, was used to jointly fit a number of published projections of the neutrino-mode CCQE and SPP event selections.
A binned $\chi^2$ test between the generated (and reweighted) events and the released data was extremised to find the best fit parameter values.
Event rate distributions, such as BNL CC1\tpip $Q^2$, \figpreref\ref{fig:bnlcc1pipq2}, were included in a shape-only way.
Cross sections, such as ANL CCQE flux-unfolded \tsenu, \figpreref\ref{fig:anlccqeenu}, were also used in the fit.
A consistent goodness of fit test, such as $\chi^2$ per number of degrees of freedom, is difficult to define because the bin-to-bin covariances were not provided with the data.
The pre- and post-fit parameter values are presented in \tablepreref\ref{table:ANLBNLFitResults}.
The fit converged and the best fit values of \tmaqe\ and \tmares\ were within the uncertainties of the nominal NuWro values.
The best fit for \tcaf\ was found to be lower than in \cite{GraczykSobczykFits_09}, however, this fit allowed the non-resonant background contribution to vary through \tnrbkg\ and included a different subset of the available data. A similar value of \tcaf\ was found in \cite{valencia_deltaspp_10} .
As part of the fit, MINUIT\,\cite{minuitref_94} calculates an approximate parameter error matrix, the corresponding  correlation matrix is shown in \figpreref\ref{fig:covmat}.
This preliminary tune did not include comparison to any hadronic mass distributions---where the effect of \tcaf\ and \tnrbkg\ might be separable.
Future tunes will include more data sets and further validation.

Event reweighting has been added to the NuWro event generator.
This enables more sophisticated investigations into the compatibility of models with data, as well as determination of well motivated, correlated model uncertainties for use in neutrino scattering analyses.

\end{document}